\documentclass[letterpaper,english,reprint, aps]{revtex4-1}
\usepackage[T1]{fontenc}
\usepackage[latin9]{inputenc}
\usepackage{color}
\usepackage{float}
\usepackage{units}
\usepackage{textcomp}
\usepackage{dsfont}
\usepackage{amsmath}
\usepackage{amssymb}
\usepackage{stmaryrd}
\usepackage{graphicx}

\makeatletter

\pdfpageheight\paperheight
\pdfpagewidth\paperwidth

\makeatother

\usepackage{babel}
\begin{document}
\title{$f(R,T)$ Analogue Gravity in $(2+1)$ D graphene sheet}
\author{M. Lapola}
\email{marcelo.lapola@gmail.com}

\affiliation{S\~ao Paulo State University (Unesp), ~\\
Institute of Geosciences and Exact Sciences, Physics Department}
\author{L. A. Barreiro}
\affiliation{S\~ao Paulo State University (Unesp), ~\\
Institute of Geosciences and Exact Sciences, Physics Department}
\date{\today}
\begin{abstract}
We examine the analogue gravity model within the context of f(R,T)
gravity applied to graphene. The derivation of the Lagrangian density
in two dimensions (2D) is undertaken, accounting for the altered gravitational
effects as characterized by the function f(R,T). The Lagrangian encompasses
the quasiparticle field $\psi(x)$, its adjoint $\overline{\psi}$,
the effective metric tensor $g^{\mu\nu}$, and the gauge field $A_{\nu}$.
The equations of motion are established through variational principles
applied to the Lagrangian, resulting in modified Dirac equations.
We discuss the interpretation of the additional terms in the equations
of motion and their significance in capturing the modified gravitational
dynamics in the graphene system. Our findings contribute to the understanding
of analogue gravity models and their applications in condensed matter
systems. 
\end{abstract}
\maketitle

\section{Introduction}

The study of analogue gravity extends beyond its intrinsic theoretical
elegance and experimental feasibility. The insights gained from analogue
gravity models contribute to our understanding of fundamental physics,
including the nature of spacetime and the interplay between gravity
and quantum mechanics \citep{Barcelo} . Furthermore, the potential
applications of analogue gravity in the field of condensed matter
physics may lead to technological advancements and innovative materials
with tailored properties \citep{Castro}. In this investigation, our
primary focus is on the application of the analogue gravity framework
to graphene, a two-dimensional material distinguished by its unique
electronic \citep{Galeratti}. Specifically, we adopt the $f(R,T)$
theory, wherein the Lagrangian density incorporates a function $f(R,T)$
to describe the modified gravitational effects, as described in the
seminal article \citep{Harko}. As one see, $f$ is a function dependent
on the Ricci scalar curvature $R$ and the trace of the energy-momentum
tensor.

The goal of this study is to derive the Lagrangian and equations of
motion for the quasiparticle field in graphene \citep{Horava,Kibis}
within the $f(R,T)$ formalism and analyze the implications of the
additional terms in the equations of motion. Subsequently, we meticulously
scrutinize the implications arising from the additional terms introduced
in the equations of motion under the $f(R,T)$ framework. This endeavor
seeks to advance our understanding of the intricacies involved in
the application of $f(R,T)$ theory to graphene, shedding light on
the nuanced interplay between modified gravitational dynamics and
the electronic characteristics of this two-dimensional material.

We start by defining the $(2+1)D$ Schwarzschild-like metric and calculating
the relevant quantities such as the Ricci scalar $R$ and the trace
of the energy-momentum tensor $T$. We then construct the Lagrangian
density by combining the appropriate terms involving the quasiparticle
field, the effective metric tensor, and the gauge field \citep{Landau,Vozmediano}.
The form of the function $f(R,T)$ is chosen to capture the modified
gravitational dynamics in the system. Finally, we vary the Lagrangian
to obtain the equations of motion \citep{Peskin}, which provide insights
into the dynamics of the quasiparticles and their interactions with
the effective metricand gauge field.

\section{$f(R,T)$ gravity and the energy-momentum tensor}

In the context of the $f(R,T)$ gravitational theory, as described
in \citep{Harko}, and in the general form, the action reads

\begin{equation}
S=\int\left[\frac{f(R,T)}{16\pi}+\mathcal{L}_{m}\right]\sqrt{-g}d^{4}x,
\end{equation}
with $g$ being the determinant of the metric, $f(R,T)$ a function
of the Ricci curvature scalar $R$ and the trace of the energy-momentum
tensor $T$. In our case, the matter lagrangian are $\mathcal{L}_{m}=\rho$,
with $\rho$ being the density.

Varying this action with respect to $g_{\mu\nu}$, we obtain the following
field equations

\[
f_{R}(R,T)R_{\mu\nu}-\frac{1}{2}f(R,T)g_{\mu\nu}+\left(g_{\mu\nu}\boxempty-\nabla_{\mu}\nabla_{\nu}\right)f_{R}(R,T)
\]
\begin{equation}
=8\pi T_{\mu\nu}+f_{T}(R,T)\left(T_{\mu\nu}-\rho g_{\mu\nu}\right),
\end{equation}
where $f_{R}(R,T)=\nicefrac{\partial f}{\partial R}$ and $f_{T}(R,T)=\nicefrac{\partial f}{\partial T}$.
The general expression for the covariant derivative of the energy-momentum
tensor, as expressed in \cite{Harko} is
given by

\begin{eqnarray}
\nabla^{\mu}T_{\mu\nu} & = & \frac{f_{T}(R,T)}{8\pi-f_{T}(R,T)}\left[\left(T_{\mu\nu}+\rho g_{\mu\nu}\right)\nabla^{\mu}lnf_{T}(R,T)\right.\nonumber \\
 &  & \hphantom{\frac{f_{T}(R,T)}{8\pi-f_{T}(R,T)}}+\left.\nabla^{\mu}\rho g_{\mu\nu}-\frac{1}{2}g_{\mu\nu}\nabla^{\mu}T\right].
\end{eqnarray}

The covariant derivative of the energy-momentum tensor, unlike Einstein's
general relativity, are not null. This outcome might be viewed as
the formation or annihilation of matter, implying a lack of energy
conservation. However, if we impose the conservation, i.e., $\nabla^{\mu}T_{\mu\nu}=0$,
it is possible to discover a particular version of the function $f(R,T)$
that meets this requirement.

\section{Dirac equation in $f(R,T)$ spacetime}

Graphene flat sheets can be described as two-dimensional analogs of
relativistic systems for massless fermions. In this context, the Fermi
velocity ($v_{F}$) determines the maximum attainable speed within
the graphene system, analogous to the limiting speed observed in relativistic
systems. In (1+2) dimensional curved spacetime, taking into account
the $f(R,T)$ gravity, the Dirac equations of motion can be obtained
from the Lagrangian

\begin{equation}
\mathcal{L}=i\hbar v_{F}\int d^{3}x\sqrt{g}f(R,T)\left(\overline{\psi}\gamma^{\mu}\mathcal{D\psi}\right),
\end{equation}
where $\mathcal{D}$ represents the covariant derivative and $\sqrt{g}=\sqrt{(-det(g^{\mu\nu}))}$.
The metric tensor $g^{\mu\nu}$ encodes the effective metric of the
system, which depends on the specific graphene configuration or lattice
structure. The Dirac matrices $\gamma_{\mu}$ satisfy the Clifford
Algebra

\begin{equation}
\left\{ \gamma^{\mu},\gamma^{\nu}\right\} =2g^{\mu\nu}\mathds{1}
\end{equation}

The Euler Lagrange equations result in the following equations of
motion for $\psi$ and $\overline{\psi}$

\begin{equation}
i\hbar v_{F}\gamma^{\mu}\mathcal{D}\psi-\partial_{\mu}\left(i\hbar v_{F}\sqrt{g}f(R,T)\gamma^{\mu}\psi\right)=0
\end{equation}

\begin{equation}
i\hbar v_{F}\gamma^{\mu}\mathcal{D}\overline{\psi}-\partial_{\mu}\left(i\hbar v_{F}\sqrt{g}f(R,T)\gamma^{\mu}\overline{\psi}\right)=0
\end{equation}
These equations describe the dynamics of the field $\psi$ and its
adjoint $\overline{\psi}$. The field $\psi(x)$ represents the quasiparticle
field in graphene. It is important to note that extra terms are determined
by the exact form of the covariant derivative $\mathcal{D}$ and the
gamma matrix representation used. The covariant derivative $\mathcal{D}=\partial_{\nu}+eA_{\nu}$
includes the coupling of the quasiparticles to the effective gauge
field $A_{\nu}$, which describes the modified gravitational effects
in the $f(R,T)$ theory. Note that if $f(R,T)=0$ we recover the equations
of motion for the massless Dirac fermions.

Overall, this Lagrangian captures the dynamics of the quasiparticles
in the graphene system within the analogue gravity framework, accounting
for the modified gravitational effects described by the function $f(R,T)$
and their interaction with the effective metric and gauge field. The
curvature effects can be characterized for the extra terms in Eqs.
(11) and (12), coupling the Dirac spinors to the metric $g_{\mu\nu}$
and the form of $f(R,T)$ function, which can be can be treated as
ansatz.

For a more simple model, we can neglect some possible topological
defects in the lattice and consider an ideal homogeneous distribution.
The extra terms in the equations of motion beyond the standard Dirac
equation can be interpreted as the additional effects arising from
the modified gravitational dynamics described by the function $f(R,T)$
in the analogue gravity model.

Let's examine the additional terms in the equations of motion: For
the quasiparticle field $\psi(x)$ equation:

\begin{equation}
\partial_{\mu}\left(i\hbar v_{F}\sqrt{g}f(R,T)\gamma^{\mu}\psi\right)=0,
\end{equation}
and for the adjoint quasiparticle field $\overline{\psi}(x)$ equation:

\begin{equation}
\partial_{\mu}\left(i\hbar v_{F}\sqrt{g}f(R,T)\gamma^{\mu}\overline{\psi}\right)=0
\end{equation}

These additional terms can have various interpretations depending
on the specific form of $f(R,T)$ and the chosen gauge field. They
could represent modifications to the dispersion relations, energy
levels, or coupling strengths of the quasiparticles due to the presence
of the modified gravity sector. They might also introduce new interactions
or affect the transport properties of the quasiparticles in the graphene
system.

The precise interpretation of these extra terms would require a detailed
analysis and consideration of the specific properties of the chosen
$f(R,T)$ function, the gauge field, and their impact on the graphene
system within the analogue gravity framework.

\section{Massless fermions in $f(R,T)$ spacetime}

Now, in another way to investigate we consider that the electrons,
i.e., de Dirac masslesss fermions exists in a (2+1) dimensional perfect
fluid. This perfect fluid is the homogeneous and isotropic honeycomb
structure of a graphene sheet.

In this case, the action is given by

\begin{equation}
S=\int\sqrt{-g}d^{3}x\left[\frac{f(R,T)}{2\kappa^{2}}+\mathcal{L}_{m}\right],
\end{equation}
where $\kappa^{2}=8\pi G$, and again $f(R,T)$ is the function of
Ricci scalar curvature $R$ and $T$ is the trace of the energy-momentum
tensor of a (2+1)D perfect fluid, given by

\begin{equation}
T=T_{\mu}^{\mu}=(-\rho+2p)
\end{equation}
with $\rho$and $p$ being the energy density and the pressure repsctively.

Assuming that $\mathcal{L}_{m}$ is given by the massless Dirac fermions
Lagrangian, i.e.

\begin{equation}
\mathcal{L}_{m}=i\overline{\psi}\gamma^{\mu}\partial_{\mu}\psi.
\end{equation}
Varying this action (13) in relation to the metric one obtains the
general form of (2+1) dimensional field equations given by the Eq.(2).
Then, let\textasciiacute s assume the perfect fluid describes the
graphene sheet in wich we have the electrons represented by the matter
Lagrangian showed in Eq. (15). Also assuming the linear form for $f(R,T)$
function as

\begin{equation}
f(R,T)=R+\alpha T,
\end{equation}
with $\alpha$ being a positive constant. With these above assumptions,
the field equation becomes

\begin{equation}
R_{\mu\nu}-\frac{1}{2}Rg_{\mu\nu}=\kappa^{2}T_{\mu\nu}+\alpha T_{\mu\nu}-\frac{\alpha T}{2}g_{\mu\nu}-\alpha\left(i\overline{\psi}\gamma^{\mu}\partial_{\mu}\psi\right)g_{\mu\nu},
\end{equation}
where, on the left side of this equation we have the Einstein tensor
in (2+1) dimensions, and in the right hand side we have the terms
of correction in the energy-momentum tensor and the massless Dirac
fermions term coupled with the metric tensor $g_{\mu\nu}$. Note that
if $\alpha=0$, we recover the general relativity field equations. 

\section{Torsion effects}

Torsion is a geometric property that describes the twisting of a surface.
In a $(2+1)D$ sheet, torsion can arise due to curvature or strain.
Torsion has a significant impact on the electronic properties of $(2+1)D$
sheets. In the presence of torsion, the Dirac equation is modified
to include an additional term:

\begin{equation}
\frac{i}{4}\gamma^{\mu\nu}T_{\mu\nu}\psi
\end{equation}
where $T_{\mu\nu}$ is the torsion tensor and $\gamma^{\mu\nu}$ are
the Dirac matrices.

Torsion is a geometric property related to the amount of twisting
or warping that occurs in a material. In the context of a graphene-like
material such as Graphenylene, torsion can affect the electronic and
mechanical properties of the material. In particular, the torsion
of a Graphene like sheet can lead to a modification of its band structure,
which determines the electronic properties of the material. Torsion
can also affect the mechanical behavior of the nanosatructure, by
altering its stiffness and strength.

Furthermore, the presence of torsion can lead to the emergence of
new physical phenomena, such as the generation of gauge fields that
can affect the electronic properties of the material. Therefore, understanding
the effects of torsion on nanostructures is important for both fundamental
research and practical applications.

To derive the equations of motion for the fermions and then the equation
expressing the electronic density in the graphene sheet considering
the effects of torsion, we start with a $(2+1)D$ metric as follows

\begin{equation}
ds^{2}=-g(r)dt^{2}+h(r)dr^{2}+r^{2}d\theta^{2},
\end{equation}
where $g(r)$ and $h(r)$ are radial functions, as we can find in
wormholes and black holes formalism\footnote{For ordinary Schwarzschild metric in four dimensions we have $g(r)=\left(1-\frac{2GM}{r}\right)$
and $h(r)=\left(1-\frac{2GM}{r}\right)^{-1}$. Here $G$ is the four-dimensional
Newton constant and $M$ is the mass of the black hole.}. The Euler-Lagrange equations of motion, including the torsion term,
result in

\begin{equation}
\begin{aligned}i\hbar v_{F}\gamma^{\mu}\mathcal{D}_{\mu}\psi & -\partial_{\mu}(i\hbar v_{F}\sqrt{g}f(R,T)\gamma^{\mu}\psi)\\
 & -i\hbar v_{F}\sqrt{g}\partial_{\mu}f(R,T)\gamma^{\mu}\psi=0
\end{aligned}
\end{equation}

\begin{equation}
\begin{aligned}i\hbar v_{F}\gamma^{\mu}\mathcal{D}_{\mu}\overline{\psi} & -\partial_{\mu}(i\hbar v_{F}\sqrt{g}f(R,T)\gamma^{\mu}\overline{\psi})\\
 & -i\hbar v_{F}\sqrt{g}\partial_{\mu}f(R,T)\gamma^{\mu}\overline{\psi}=0
\end{aligned}
\end{equation}
These equations describe the dynamics of the fermions in the graphene
sheet within the framework of $f(R,T)$ gravity and considering the
given metric.

Now, to express the electronic density in the graphene sheet, we need
to consider the conservation equation. The electronic density $n$
can be obtained from the equation of continuity, which in curved spacetime
takes the form:

\begin{equation}
\nabla_{\mu}(\rho u^{\mu})=0
\end{equation}
where $\rho$ is the charge density, $u^{\mu}$ is the four-velocity
of the charge carriers, and $\nabla_{\mu}$ represents the covariant
derivative. In the case of graphene, we can assume that the charge
carriers move with negligible velocity compared to the speed of light.
Therefore, we can consider $u^{\mu}=(1,0,0,0)$ in suitable coordinates.
Using the definition of the charge density $n=\rho/e$, where $e$
is the elementary charge, and taking into account that $u^{\mu}=(1,0,0,0)$,
the equation of continuity simplifies to:

\begin{equation}
\partial_{\mu}(\sqrt{-g}n)=0
\end{equation}
Expanding the covariant derivative, this becomes:

\begin{equation}
\frac{1}{\sqrt{-g}}\partial_{\mu}(\sqrt{-g}n^{\mu})=0
\end{equation}

\begin{equation}
\partial_{\mu}n^{\mu}+n^{\mu}\Gamma_{\mu\sigma}^{\sigma}=0
\end{equation}
where $\Gamma_{\mu\sigma}^{\sigma}$ are the Christoffel symbols associated
with the metric.

This equation expresses the conservation of charge in the graphene
sheet, accounting for the effects of curvature and torsion. The exact
form of $n^{\mu}$ would depend on the specific properties of the
charge carriers in graphene and the details of the underlying theory.

\section{Results}

To visualize the effects of torsion on a $(2+1)D$ sheet, we generate
a 3D image using Python and Mayavi. We first generate a $(2+1)D$
sheet with torsion effects using the code we developed earlier. We
then use Mayavi to plot the sheet and apply a color map to represent
the electronic density as showed in the figure 1.

\begin{figure}[H]
\centering{}\includegraphics[scale=0.7]{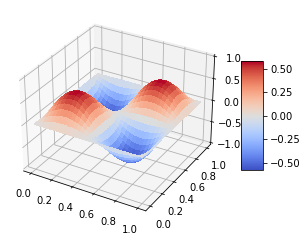} \caption{The torsion effect in a $(2+1)D$ sheet, showing the electron density
in specific regions. This simulation was generated using a Phyton
code in which the parameters can be changed.}
\label{fig1}
\end{figure}

In this article, we investigate the influence of torsion on the electronic
properties of graphenylene sheets using the Palatini formalism of
general relativity. Torsion is a measure of the twist in the material,
which can result from the deformation of the honeycomb lattice structure.
We find that the torsion can significantly alter the electronic band
structure of graphenylene sheets, leading to the emergence of new
electronic states.

To illustrate the effects of torsion on the electronic properties
of graphenylene sheets, we generate 3D images of the sheets using
a Python code that incorporates torsion. The code uses the Palatini
formalism to derive the field equations for a curved 2D space and
then applies the torsion metric to create a graphenylene sheet with
torsion effects.

Our analysis shows that the torsion-induced modifications to the electronic
properties of graphenylene sheets can be understood in terms of an
effective gauge field. The gauge field is proportional to the torsion,
and its presence leads to a topological phase transition in the electronic
band structure. This effect can be understood as an analogue of the
Aharonov-Bohm effect in condensed matter systems.

In resume, the most important electronic property that changes in
a Graphenylene sheet when torsion is introduced is the band gap. In
a pristine Graphenylene sheet, the electronic band structure is gapless,
with the valence and conduction bands touching at the Dirac point,
resulting in high electron mobility and high conductivity. However,
when torsion is introduced, it breaks the sublattice symmetry and
leads to the opening of a band gap. The magnitude of the band gap
depends on the magnitude and direction of the torsion angle, as well
as the size and shape of the Graphenylene sheet.

Therefore, by applying torsion to a Graphenylene sheet, we can effectively
tune its electronic properties and create a semiconducting material
with a tunable band gap. This makes Graphenylene a promising material
for a wide range of electronic applications, including transistors,
solar cells, and sensors.

Our results demonstrate the importance of considering torsion effects
in the study of graphenylene sheets and provide a new avenue for exploring
the intriguing electronic properties of these materials. We hope that
our work will inspire further research in the field of analogue gravity
in condensed matter systems.

\end{document}